# THE CONTROL SYSTEM MODELING LANGUAGE

K. Zagar, M. Plesko, M. Sekoranja, G. Tkacik, A. Vodovnik
J. Stefan Institute, SI-1000 Ljubljana, Slovenia

Abstract

The well-known Unified Modeling Language (UML) describes software entities, such as interfaces, classes, operations and attributes, as well as relationships among them, e.g. inheritance, containment and dependency. The power of UML lies in Computer Aided Software Engineering (CASE) tools such as Rational Rose, which are also capable of generating software structures from visual object definitions and relations. UML also allows add-ons that define specific structures and patterns in order to steer and automate the design process.

We have developed an add-on called Control System Modeling Language (CSML). It introduces entities and relationships that we know from control systems, such as "property" representing a single controllable point/channel, or an "event" specifying that a device is capable of notifying its clients through events. Entities can also possess CSML-specific characteristics, such as physical units and valid ranges for input parameters.

CSML is independent of any specific language or technology and generic such that any control system can be described with it. Simple transformation scripts map CSML defined structures to APIs and tools such as EPICS, CDEV, SCADA, Abeans, BACI and generate the appropriate database or source files.

Advantages of control system development with CSML are discussed on a concrete example of a bending magnet's power supply in a synchrotron accelerator.

## 1 INTRODUCTION

### 1.1 Unified Modeling Language (UML)

The Unified Modeling Language (UML, [1]) is an open industry standard for specifying, visualizing, constructing, and documenting the artifacts of software systems, as well as for business modeling and other non-software systems. Now it is under supervision of the Object Management Group, but it emerged in the mid- '90s as a confluence of several different software modeling methodologies.

### 1.2 Computer Aided Software Engineering Tools (CASE)

Nowadays, several tools are available on the market to aid software engineering. Typically, these tools allow the software developers to gather requirements, visually enumerate and define the building blocks, and allow forward and reverse engineering of the software system. They typically employ the standard UML notation.

We have used Rational Rose [2] as our CASE tool. In particular, we have written an add-on extension for Rose which generates XML representation of a CSML description, and facilitates further transformation into other software artifacts using XSL [6] (for details, see section on generators).

### 1.2 Extensible Markup Language (XML)

Extensible Markup Language (XML, [3]) provides a standard way to create structured textual documents. The tree-like structure of a document is denoted using markups (tags) embedded in the text. Today, XML is becoming a widely accepted method of representing data of all kinds, and is suitable for data transfer as well as persistent storage.

Major advantage of XML is its ability to be transformed into another XML document with a different schema, or into a plain text or program source code through the use of Extensible Stylesheet Language (XSL, [4]).

As such, XML is the ideal storage for control system's data (the configuration database), as well as meta-data (the description of the control system's structures and their relationships).

## 2 THE CONTROL SYSTEM MODELING LANGUAGE

UML was designed to be applicable to a wide class of software engineering applications, ranging from construction of control system software to modeling of business processes. Control System Modeling Language ([5]), however, is a specialized dialect of the Unified Modeling Language. As such, it uses the same terminology and notation as UML. The following UML terms are commonly used in CSML:

1. A **class** encapsulates data (**attributes**) and behavior (**operations**). A *power supply* is an example of a class, whose attribute might be the *current* it produces, and whose operations might be *On* and *Off*.
2. An **object** is a concrete instance of a class. For example, the power supply of the 4th bending magnet in a given synchrotron is an *object* of the *power supply* class.
3. **Associations** describe relationships between entities in the model.

Each of these entities can have several properties assigned to it that define them. For example, every class has a name, and every attribute has at least a name and a type. One of the standard properties is the **stereotype**, which is very useful for further classification of entities. For example, some classes might describe devices, whereas others could describe state machines.

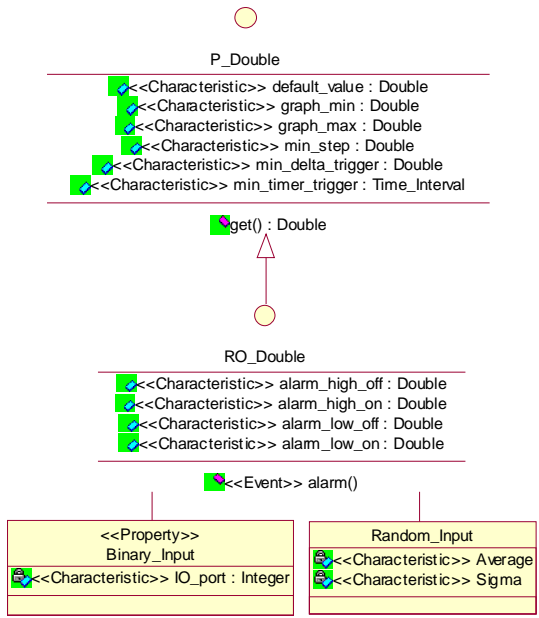

**Figure 1: Interface to a read-only (RO) property of type Double, and two classes implementing the same interface. The `Binary_Input` class (an equivalent of EPICS' binary input) is implemented by reading the desired value from hardware (e.g., an ADC converter), whereas `Random_Input` just produces Gaussian random values with a given average and standard deviation (simulation).**

A control system may be described using CSML regardless of whether it is based on EPICS, BACI or other infrastructure. However, a different set of generators must be provided for every target infrastructure to enable automatic generation of the artifacts.

## 2.1 CSML Concepts

All **basic types** in CSML must be declared before they are used. In EPICS, an equivalent of the basic type is the *value type*.

**Characteristics** are attributes of a basic type whose purpose is to characterize the entity in which they are contained. Because entity's character is static and immutable at run-time, characteristics are usually stored in the configuration database. The use of characteristics is illustrated in Figure 1. Characteristics are very similar to EPICS' *fields*.

Especially in control systems it is crucial to be capable of notifying interested parties whenever a change occurs, for example to raise alarms when unsafe conditions appear. Such notifications are accomplished through the use of **events**. Events are equivalent to EPICS' *monitors* and *alarms*.

**State machines** model the state of devices supervised by the control system. The state represented by a state machine is influenced by values of attributes, CSML properties, events and operations.

Individual observables in a control system, such as a given temperature sensor's readout, could be modeled as simple UML attributes of a basic type, e.g., a double, measuring the temperature in Kelvin. A better approach is to introduce more complex types, **properties**, which are stereotyped classes, and may provide support for characteristics and events, as well as operations other than *get* and *set* (Figure 1). Properties are equivalent to EPICS' *records*.

A **device** represents a physical device, e.g., a vacuum gauge, a radio-frequency cavity or a power supply. In CSML, a device is also a stereotyped class. Usually, devices are composed of properties. For example of a device, see Figure 2.

## 3 GENERATORS

We have designed an XML schema ([5]), which is particularly suitable for storing control system descriptions in XML format. Once the control system is described in XML, it is possible to use Extensible Style-sheet Language (XSL, [4]) transformations as generators that produce:

- EPICS template and State Notation Language (SNL) files.

- Schemas for XML-based configuration database.
- User interfaces for manipulating individual devices.

In particular, we have implemented these generators:

1. Java source code for Java Beans representing devices. These beans (Abeans) allow visual composition of user interfaces for end-users of the control system.
2. C++ source code for CORBA device servers, linking the device drivers with the CORBA middleware.
3. Delphi source code for HTTP protocol listeners that exposed COM objects via Web Services (SOAP); developed for a commercial company.

We have found that XSLT is excellent tool for producing XML documents. However, generating text files is more painful, as one is forced to balance between the legibility of the XSL transform and the resulting text file. To overcome this disadvantage, we have decided to construct a language with all the XSL's strengths (processing of XML files, existing XSL-related standards, …), while trying to correct the lack of control over spacing, and add the ability to insert custom user code in the generated output, and retain these modifications after subsequent regenerations of the output files. We called this language Extensible Program Generator Language (XPGL), and the first program generator using it is currently being implemented.

## 4 CONCLUSION

CSML allows definition of the control system's structure in a single place. All dependant artifacts, such as program source code, configuration database schema and technical documentation can be generated from that single source.

Because CSML is essentially standard UML, it also serves as the blueprint for the control system. Not only is it standardized, it also offers a visual, easy to comprehend notation of system's building blocks and their relationships.

CSML's one-to-one mapping to its XML representation offers a starting point from which other artifacts may be generated using standard tools and leverages widely available knowledge of XSL.

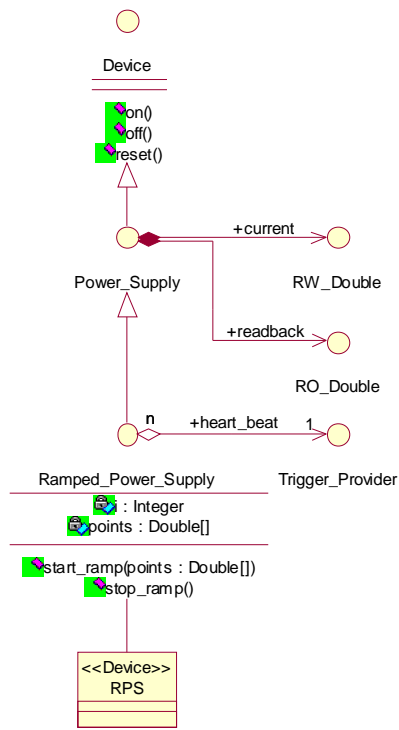

**Figure 2: A ramped power supply (RPS) device. A ramped power supply implements a Ramped_Power_Supply interface, which inherits all attributes and operations from a Power_Supply, but adds a reference to another device that provides it with heartbeat required for synchronized ramping. A regular power supply already contains a writable double property for setting the current, and a read-only property for reading out the actual current.**

## REFERENCES

[1] Object Management Group (OMG), "Unified Modeling Language (UML) 1.4 specification", September 2001. Available at OMG web site, http://www.omg.org.
[2] Rational, "Rational Rose", http://www.rational.com/rose.
[3] The World Wide Web Consortium, "Extensible Markup Language (XML) 1.0 (Second Edition)", October 2000, http://www.w3.org/XML.
[4] The World Wide Web Consortium, "Extensible Stylesheet Language (XSL)", May 2001, http://www.w3.org/Style/XSL/.
[5] K. Zagar, IJS, "Control System Modeling Language Specification", October 2001. Available at http://kgb.ijs.si.
[6] K. Zagar, IJS, "CSML Add-on for Rational Rose", http://kgb.ijs.si.
[7] K. Zagar, IJS, "The Control System Modeling Language" (full version of this article), http://kgb.ijs.si.